
\input phyzzx

\def\ltsima{$\; \buildrel < \over \sim \;$}
\def\simlt{\lower.5ex\hbox{\ltsima}}
\def\gtsima{$\; \buildrel > \over \sim \;$}
\def\simgt{\lower.5ex\hbox{\gtsima}}

\PHYSREV
\singlespace

\def\skipfig#1#2#3{\midinsert
                   \vskip #1 truecm
                   \hbox{\hskip 1. truecm
                         \vbox{\hsize=13.1 truecm
                               \noindent {\tenrm {\bf Figure #2:} #3} } }
                   \endinsert}
\def \figureinsert#1;#2;{\midinsert\noindent\ \unskip\hskip-\hoffset
\special{overlay #1}\vskip11truecm\input #2 \endinsert}
\def\one{$^1$}
\def\two{$^2$}
\def\three{$^3$}
\def\four{$^4$}
\def\five{$^5$}
\def\onetwo{$^{1,2}$}
\def\twofour{$^{2,4}$}
\def\twofive{$^{2,5}$}

\pubnum{}
\pubnum{UCRL-JC-11113606}
\pubtype{}
\date{}
\titlepage
\singlespace
\title{\bf The First Data from the MACHO Experiment\foot{
Presented at the Texas/PASCOS Symposium,
Berkeley, CA, Dec. 1992; to appear in the Proceedings.}}
\author{David P. Bennett\onetwo}

\vskip .2in
\centerline{in collaboration with}
\centerline{C.~Akerlof\two\llap, C.~Alcock\one\llap, R.~Allsman\one\llap,
T.~Axelrod\one\llap,
K.H.~Cook\rlap,\onetwo\ K.~Freeman\three\llap,}
\centerline{K.~Griest\rlap,\twofour\ S.~Marshall\rlap,\twofive\
H.-S.~Park\one\llap, S.~Perlmutter\two\llap,
B.~Peterson\three\llap,}
\centerline{ P.~Quinn\three\llap, A.~Rodgers\three\llap,
C.W.~Stubbs\rlap,\twofive\ and W.~Sutherland\two}
\vskip .13in
\centerline{\one Lawrence Livermore National Laboratory, Livermore, CA 94550}
\centerline{\two Center for Particle Astrophysics, Univ. of California,
Berkeley, CA 94720}
\centerline{\three Mt. Stromlo and Siding Spring Observatories, ACT 2611 Weston
Creek, Australia}
\centerline{\four Department of Physics, Univ. of California San Diego,
La Jolla, CA  92093}
\centerline{\five Department of Physics, Univ. of California, Santa Barbara, CA
93106}

\abstract
\singlespace
MAssive Compact Halo Objects such as
brown dwarfs, Jupiters, and black holes are prime candidates to comprise the
dark halo of our galaxy. Paczynski noted that objects (dubbed MACHOs)
with masses in the range $10^{-6}M_\odot < M \simlt 100 M_\odot$.
can be detected via gravitational microlensing of stars in the Magellanic
Clouds with the caveat that only about one in $10^6$ stars will be lensed
at any given time. Our group has recently begun a search
for microlensing using a refurbished 1.27 meter telescope
at the Mount Stromlo Observatory in Australia. Since the summer of 1992, we
have been imaging up to $10^7$ stars a night in the Large Magellanic Cloud
using our
large format two-color $3.4\times 10^7$ pixel CCD camera. Here I report on
our first results based on an analysis of $\sim 10^6$ of these stars. Although
this is not enough data to make definitive statements about the nature of
the dark matter, we are able to conclude that the rate of variable star
background events is not larger than the expected MACHO signal.

\singlespace
\chapter{Introduction}

   The nature of the dark matter which dominates the mass of galaxies
is arguably the most important unsolved problem in astronomy. The
most fasionable idea about the nature of this dark matter is that it
is made up of exotic elementary particles such as WIMP's or axions that
have never been observed. The discovery of
anisotropies in the Cosmic Microwave Background Radiation by the
COBE-DMR team
  \Ref\Smoot{G.~F.~Smoot, \etal, {\sl Astrophys. J.} {\bf 396}, L1 (1992).}
has been widely hailed by some prominant scientists and
in the popular press as providing strong support for the idea that the
dark matter is composed primarily of exotic elementary particles.
In fact, although the COBE-DMR detection does provide strong support for the
idea that the structure in the universe is a result of gravitational
amplification of small initial perturbations, it does not support the
simplest model based on exotic dark matter (the standard Cold Dark Matter
model). The {\it true} state of affairs was revealed at the
Cosmic Microwave Background Workshop held at LBL immediately prior to this
meeting. After hearing numerous talks about models with ``tilted spectra"
and two types of exotic dark matter all contrived to fit the COBE data,
the participants were asked to vote on which theory stood the best
chance of being correct. The winner with an overwelming plurality
of the vote was a ``baryons only" cosmology with no exotic dark matter, so
the case for non-baryonic matter is apparently not as strong as many have
been led to believe. In the remainder of this talk, I will argue that
we will soon have a better way of determining the composition of the
universe, as dark matter search experiments will soon be generating
definitive constraints on the nature of the dark matter.

If the galactic dark matter is composed of baryons, then it cannot be made of
normal stars, dust, or gas, which would readily be detected. Instead, it
would have to be in the form of brown dwarfs, ``Jupiters\rlap,"
white dwarfs, neutron stars, or black hole
stellar remnants. These objects can all be detected by the
technique of gravitational microlensing, and have come to be known as MACHOs
(for MAssive Compact Halo Objects).

Gravitational microlensing refers to lensing in the case where their
is significant amplification of the source, but the lensing angle is too
small to be observed. Paczynski
   \REFS\pacz{B.~Paczynski, {\sl Astrophys. J.} {\bf 304}, 1 (1986).}
   \REFSCON\kg{K.~Griest, {\sl Astrophys. J.} {\bf 366}, 412 (1991).}\refsend
has shown that the ``optical depth" for microlensing by the dark halo
of our galaxy is $\sim 5\times 10^{-7}$,
so that at any given time about 1 star in $2\times 10^6$ will be microlensed
with amplification by a factor of 1.34 or larger. (The typical angular
separation of the images is $\simlt 0.001\,$arc second, so they cannot be
resolved.) The fundamental unit of
length for a microlensing is the Einstein ring radius which is given by
$$ R=2\,\left({GMx(L-x)\over c^2 L}\right)^{1/2} $$
where $L$ is the distance to the source star (in our case in the Magellanic
Clouds) and $x$ is the distance to the lensing object in the halo. If
$R$ is not too small, the source star can be assumed to be a point
light source, and the microlensing amplification depends only on the
dimensionless impact parameter ($u\equiv b/R$):
$$ A(u) ={u^2+2\over u\sqrt{u^2+4}} $$
$u=1$ corresponds to an amplification of 1.34.
For source stars in the Magellanic clouds, the point-like approximation
fails only when the mass of the lensing objects is less than
$M \simlt 10^{-7} M_\odot$.

The time scale of a microlensing event is given by the timescale for the
lensing object to traverse a distance equal to the impact parameter $b$
which yields
$$ \Delta t = 7\,{\rm days}\,\sqrt{M\over 10^{-2} M_\odot} $$
for a typical lensing object located at a distance of $10\,$kpc moving
at a velocity of 200 km/sec.

The microlensing light curve will be distinguishable from the background
of variable stars in several important ways: they are achromatic,
time-symmetric, and non-repeating. These signatures should allow us
to distinguish most (and hopefully all) of the variable stars from
the microlensing events. In addition, the microlensing light curve
is described by just 3 parameters: the maximum amplification, the
time of the maximum amplification and the duration of the event.
A convincing microlensing signal would certainly require that at
least some of the detected microlensing events have a very high amplification
and a well sampled light curve.

\chapter{The Macho Telescope System}

The search for gravitational microlensing by Machos in the halo of our
galaxy requires that we survey large numbers of stars in the
Magellanic Clouds and/or Galactic Bulge on a nightly basis for several years.
To accomplish this goal, we have obtained the recently refurbished Mt. Stromlo
50" ``Great Melborne Telescope" for four years of dedicated use. Since our
survey requires a wide field of view, we have installed a system of corrector
lenses (designed by E. H. Richardson and I.Z. Lewis) which give us a
$1^\circ$ diameter field of view and
reduces the f ratio to f/3.88 (giving an image scale of 41 arcsec/mm).  In
addition, a dichroic filter within the corrector is used to split the beam into
two channels (which we have chosen to be 4500A -6300A, and 6300 - 7800A) to
give simultaneous imaging in two colors.

In order to image these large fields of view, two large CCD cameras
   \Ref\stubbs{C.~W.~Stubbs, \etal, ``A 32 Megapixel Dual Color CCD Imaging
        System\rlap," to appear in {\sl Charged Coupled Devices and Solid
        State Sensors III}, M. Blouke, ed., SPIE Proc. {\bf 1900}, (1993).}
have been fabricated (for the ``red" and the ``blue" channels) each
containing 2 x 2 mosaics of 2048 x 2048 CCDs, fabricated by Loral
Aerospace. These CCDs have two readout amplifiers per chip
and 15 micron pixels ($\simeq 0.63$" on the sky). Using all 16 readout
amplifiers simultaneously, the CCDs are read out in about 70 seconds with
a read noise of less than 10 electrons. For our typical exposure time of
300 seconds, we are able to take 8-9 exposures per hour, and we can generally
take more than 60 images on a clear night in the middle of the LMC season.

\chapter{Data Pipeline and Photometric Analysis}

Since our cameras produce data at a prodigious rate, it is highly desirable
to have a data pipeline that is both automatic and computationally
efficient.  Raw image data from the
cameras passes down a fiber optic line to 128 MBytes of buffer memory on a
VME Bus extension to a Sun Sparc IPC.  The images are then written to
magnetic disk and copied to tape. While the images are resident on disk,
our 4-processor Solbourne reduces the data to photometric measurements
using our custom built PSF fitting photometry routine, SoDOPHOT (originally
based on DOPHOT\Ref\dophot{M.~Mateo and P.~Schechter,
      in {\sl The First ESO/ST-ECF Workshop
      on Data Analysis}, ed. P. J. Grosbol, F. Murtagh, and R. H. Warmels,
      (Munich: ESO, 1989), p. 69}).
SoDOPHOT uses a template of positions and magnitudes from a good seeing image
as a starting point for its fits, and we find that the photometry obtained
with varying seeing and sky brightness is much more reproducable than with
the standard versions of DOPHOT or DAOPHOT. The photmetric output includes
a number of parameters and flags designed to help identify possible
photometric problems. The quantities output include the PSF fit $\chi^2$,
measures of cosmic ray contamination and the fraction of the PSF
that is lost to bad pixels, and a seeing-dependent crowding estimator.

Although we have established that our photometric analysis system can easily
keep up with the incoming data, the software to control the automated
photometric reduction of all our data is not yet fully opperational.
The results reported below are
based on an analysis of about 15\% of our data generated by running
SoDOPHOT outside our automatic analysis pipeline.

\chapter{First Results}

As of the end of November, 1992, we have analyzed 182 images of 3 fields
in the LMC bar. The total number of stars contained in these
fields is about 1.2 million, and the total number of two-color
photometric measurements in this data set is about 72 million. (By the end of
January, 1993, we had analyzed 419 images of 5 LMC bar fields containing
1.7 million stars.) In every image we take,
our photometry routine, SoDOPHOT, makes photometric measurements for
all the stars in that image, even those stars whose signal is below the noise
in the image being reduced. (The stellar content and relative positions
had previously been determined from the high quality ``template image".)
Thus, when the seeing is poor or the sky is bright, the photometric
errors can become large.
About 50\% of these measurements have errors of less than 15\%.

We have performed a preliminary time series analysis on this data
to search for variable stars and possible MACHO events. We have set up
cuts on a number of the SoDOPHOT ouput parameters in order to remove
suspect data. Variable star candidates are determined by selecting
stars which are not a good fit to a constant light curve. This
can be done on the full data set or a ``robust" set in which the
measurements with the largest deviations fromthe mean have been removed.
This procedure has yielded several thousand variable candiates. Periods
for some of these variables have been determined using the phase
dispersion minimization method of Stellinwerf. A few of these light curves
for short period variables are plotted in Fig. 1.
Although we rarely image the same field more
than twice a night, we are able to obtain good light curves for these stars
because there is sufficient randomness in our observing schedule.

\skipfig{17.0}{1}{Two color light curves for 2 short period variables.}

The results of our preliminary microlensing searches have been quite
encouraging. We flag microlensing candidates using optimal filters
of several different durations, and do a simultaneous microlensing curve
fit in both colors
for the candidates which passed the filters. The candidate list
generated from more than a million light curves is easily reduced
to a handful of stars by cutting out stars which are a poor fit to the
microlensing light curve, have very poor sampling or low signal to noise
near the peak, or are flagged by the SoDOPHOT output parameters as having
potential photometry problems. The result of this procedure is that, we
have no good microlensing candidates when a preliminary estimate of our
detection efficiency suggests that we should have seen 1 or 2 strong events.
Fig. 2 shows a light curve of one of our ``close calls" (star 29-3902 in
field 79). While the $\chi^2$ value for the 2-color microlens curve fit
indicates an acceptable fit, closer inspection indicates that light curve
is actually a poor fit near the peak. The 17 data points within 18 days of
the peak have a $\chi^2 = 32.4$ for (essentially) 14 degrees of freedom, so
we can exclude microlensing at better than the 99\% confidence level. We
do have a couple of other events that are consistent with microlensing,
but their signal-to-noise is too low to claim a dectection.

\skipfig{12.0}{2}{The two-color light curve of a ``near" MACHO-candidate
plotted in linear units. The unamplified magnitues ($A=1$) correspond
to $MR = 16.83$ and $MV = 17.02$. These light curves are not well fit by
a microlensing light curve near the peak.}

Our conclusion from this preliminary analysis is that while we have not
analyzed enough data to put a statistically significant limit of the
abundance of MACHOs in the Halo of our galaxy, this analysis has
clearly demonstated that the background of variable stars mimicing
microlensing events is at worst comparible to our expected signal.
Since we have not yet used all the tools at our disposal for rejecting
background events (such as location on the Color-Magnitude diagram
or the spectra of candidate stars), we believe that the variable star
background will not seriously hinder our efforts to confirm or
deny the existence of MACHOs.

\ack
We would like to thank the technical staff at the Mt. Stromlo Observatory
who have completely reworked the mechanical and optical aspects of the
Great Melbourne Telescope, and have
provided it with a state-of-the art control system.
We are also indebted to Simon Chan for his indispensable
ongoing contribution to the MACHO project.

This work was supported in part by the Center for Particle Astrophysics, a
National Science and Technology Center
operated by the University of California,
Berkeley under cooperative agreement AST-8809616. Part of this work
was performed under the auspices of
the U.S. Department of Energy at the Lawrence Livermore
National Laboratory under contract No. W-7405-Eng-48

\refout
\end